\documentclass[letterpaper,reprint,prl,superscriptaddress,nofootinbib,longbibliography]{revtex4-1}

\usepackage{amsmath,amsfonts,amssymb,amsthm, booktabs, tabularx, placeins}

\newcommand{\MP}{M_{\rm Pl}}
\begin{document}

\title{A frame independent classification of single field inflationary models}

\author{Laur J\"arv}
\affiliation{Institute of Physics, University of Tartu, W. Ostwaldi 1, 50411 Tartu, Estonia}
\author{Kristjan Kannike}
\affiliation{National Institute of Chemical Physics and Biophysics, R\"avala 10, 10143 Tallinn, Estonia}
\author{Luca Marzola}
\affiliation{Institute of Physics, University of Tartu, W. Ostwaldi 1, 50411 Tartu, Estonia}
\affiliation{National Institute of Chemical Physics and Biophysics, R\"avala 10, 10143 Tallinn, Estonia}
\author{Antonio Racioppi}
\affiliation{National Institute of Chemical Physics and Biophysics, R\"avala 10, 10143 Tallinn, Estonia}
\author{Martti Raidal}
\affiliation{Institute of Physics, University of Tartu, W. Ostwaldi 1, 50411 Tartu, Estonia}
\affiliation{National Institute of Chemical Physics and Biophysics, R\"avala 10, 10143 Tallinn, Estonia}
\author{Mihkel R\"unkla}
\affiliation{Institute of Physics, University of Tartu, W. Ostwaldi 1, 50411 Tartu, Estonia}
\author{Margus Saal}
\affiliation{Institute of Physics, University of Tartu, W. Ostwaldi 1, 50411 Tartu, Estonia}
\author{Hardi Veerm\"ae}
\affiliation{National Institute of Chemical Physics and Biophysics, R\"avala 10, 10143 Tallinn, Estonia}

\begin{abstract}
{Seemingly unrelated models of inflation that originate from different physical setups yield, in some cases, identical predictions for the currently constrained inflationary observables. In order to classify the available models, we propose to express the slow-roll parameters and the relevant observables in terms of frame and reparametrisation invariant quantities. The adopted invariant formalism makes manifest the redundancy that afflicts the current description of inflation dynamics and offers a straightforward way to identify classes of models which yield identical phenomenology. In this Letter we offer a step-to-step recipe to recast every single field inflationary model in the proposed formalism, detailing also the procedure to compute inflationary observables in terms of frame and reparametrisation invariant quantities. We hope that our results become the cornerstone of a new categorisation of viable inflationary models and open the way to a deeper understanding of the inflation mechanism.}
\end{abstract}

\maketitle

\section{Introduction} 
\label{sec:Introduction}

According to present knowledge~\cite{Ade:2015lrj}, the Universe underwent a phase of exponential expansion in the very first moments of its existence~\cite{Starobinsky:1980te,Guth:1980zm,Linde:1981mu,Albrecht:1982wi}. This period, known as \emph{inflation}, is crucial for setting the peculiar initial conditions required by the $\Lambda \mathrm{CDM}$ of cosmology~\cite{Ade:2015xua}. Although the available dedicated measurements have already shed some light on the features of inflation, the mechanism behind its dynamics still remains a mystery. 

During the last decades, the puzzle of inflation has been tackled in a multitude of approaches, encapsulated in models that originate from very different background physics. A common trait of many inflationary scenarios is that they involve scalar degrees of freedom\footnote{In this work we do not consider theories in which the inflaton is a vector degree of freedom. For a review we refer the reader to~\cite{Martin:2013tda}.} with properties beyond the Standard Model limits, and/or extend the theory of gravity. To-date, many review articles~\cite{Martin:2013tda,Olive:1989nu,Lyth:1998xn} have attempted to classify the viable models of inflation according to various criteria, most commonly by their theoretical origin.
Interestingly, despite the different starting points, there are known cases~\cite{Bezrukov:2011gp,Kallosh:2013tua,Galante:2014ifa,Kannike:2015apa,
Rinaldi:2015yoa,Kannike:2015kda,Ozkan:2015kma,Kallosh:2016gqp,Yi:2016jqr,Artymowski:2016dlz} of seemingly different models that give identical predictions for the inflationary parameters currently constrained by experiments, such as the spectral index $n_s$ and the tensor-to-scalar ratio $r$. It could then be sustained that in the current descriptions of inflation there is a redundancy which unnecessarily complicates the landscape of viable frameworks and obscures the understanding of the underlying mechanism. 

The purpose of the present Letter is to expose the origin of this redundancy and to present a clear method to sort the available inflationary models into classes of phenomenologically equivalent frameworks. To this purpose, working in the context of scalar-tensor theories of gravity, we show how the slow-roll parameters and the relevant observables can be written in a frame and reparametrisation invariant way. Our starting point are the quantities proposed originally in \cite{Jarv:2014hma, Kuusk:2016rso}, which allow for a more versatile approach than methods based purely on frame invariance \cite{Catena:2006bd,Chiba:2013mha,Postma:2014vaa,Chiba:2014sva,Burns:2016ric}.  
Our results prove unequivocally that inflationary parameters depend solely on one thing: the invariant potential. As a consequence, distinct models characterised by the same invariant potential yield, inevitably, the same phenomenological consequences. 
This conclusion allows for the sought categorisation of the known inflation frameworks in a straightforward way. The resulting classes of equivalent models expose the redundancy that afflicts the traditional formalism, paving the way for a deeper understanding of the inflationary mechanism. 

In the following, after introducing the necessary formalism, we present a detailed, cook-book level recipe to rephrase different models in a frame and reparametrisation invariant fashion under the slow-roll approximation. With this invariant formalism at hand, we then show preliminary examples of the power of the proposed categorisation by explicitly studying two sets of models that, despite different origins, yield the same inflationary phenomenology.

\section{Frame and parametrisation invariant formalism} 
\label{sec:The invariant formalism}

The action of a general scalar-tensor theory of gravity without derivative couplings and higher derivative terms is specified by four arbitrary functions of the scalar field $\Phi$~\cite{Flanagan:2004bz}: ${\mathcal A}(\Phi)$, ${\mathcal B}(\Phi)$, ${\mathcal V}(\Phi)$ and
$\sigma(\Phi)$\footnote{We have changed the notation of~\cite{Flanagan:2004bz} relabelling the function  $\alpha(\Phi)$ as  $\sigma(\Phi)$.}. Explicitly we have 
\begin{align}
	\label{Flanagan.action}
	S & = \int\mathrm{d}^4x\sqrt{-g}\Bigg[ -\frac{1}{2} \mathcal A(\Phi)\MP ^2 R +\frac{1}{2}{\mathcal B}(\Phi)g^{\mu\nu}\nabla_\mu\Phi \nabla_\nu\Phi 
	\nonumber \\&\qquad
	- {\mathcal V}(\Phi)\Bigg] 
	+ S_\mathrm{m}\left(\mathrm{e}^{2\sigma(\Phi)}g_{\mu\nu},\chi\right) \,,
\end{align}
where the Ricci scalar $R$ is determined by the metric $g_{\mu\nu}$ (the adopted signature is $(+,-,-,-)$), the reduced Planck mass is denoted by $\MP $, and $S_m$ is the action for the matter fields represented by $\chi$. 

The action~\eqref{Flanagan.action} encapsulates many inflationary models, as it allows for a non-minimal coupling of the involved scalar field $\Phi$ to curvature, a non-canonical form for its kinetic term, an arbitrary scalar potential and a possible non-minimal coupling of $\Phi$ to matter. Notice that~\eqref{Flanagan.action} also encompasses $f(R)$ theories including the Starobinsky \cite{Starobinsky:1980te} inflation because of their equivalence to scalar-tensor theories defined by the O'Hanlon action, $-(\Phi R + V(\Phi))$, where the potential $V$ is the Legendre transformation of $f$~\cite{Chiba:2003ir}.
In this Letter we will not consider single scalar actions with more complicated forms (see for instance~\cite{Kobayashi:2011nu}), however, the proposed formalism can be extended to models which generalize the action~\eqref{Flanagan.action} for multiple scalar fields~\cite{Kuusk:2015dda}.
 
The action functional \eqref{Flanagan.action} preserves its structure, up to a boundary term, under a conformal rescaling of the metric 
\begin{equation}
	\label{conformal.transformation}
		g_{\mu\nu} = \mathrm{e}^{2\bar{\gamma}(\bar{\Phi})}\bar{g}_{\mu\nu}\,,
\end{equation}
and redefinition of the scalar field, 
\begin{equation}
	\label{field.redefinition}
		\Phi = \bar{f}(\bar{\Phi}) \,,
\end{equation}
provided that the functions of the scalar field transform according to \cite{Flanagan:2004bz}:
\begin{align}
		\label{Flanagan.A}
		\bar{\mathcal{A}}(\bar{\Phi}) &= \mathrm{e}^{2\bar{\gamma}(\bar{\Phi})}
		{\mathcal A} \left( {\bar f}( {\bar \Phi})\right) \,,\\
		\label{Flanagan.V}
		\bar{{\mathcal V}}(\bar{\Phi}) &= \mathrm{e}^{4\bar{\gamma}(\bar{\Phi})} \, {\mathcal V}\left(\bar{f}(\bar{\Phi})\right) \,, \\
		\label{Flanagan.alpha}
		\bar{\sigma}(\bar{\Phi}) &= \sigma\left(\bar{f}(\bar{\Phi})\right) + \bar{\gamma}(\bar{\Phi})\, , \\
		\label{Flanagan.B}
		{\bar {\mathcal B}}({\bar \Phi}) &= \mathrm{e}^{2{\bar \gamma}({\bar \Phi})}\big[ 
		\left(\bar{f}^\prime\right)^2{\mathcal B}\left(\bar{f}(\bar{\Phi})\right) 
	 \nonumber \\	
		&	  
		-6\MP ^2\left(\bar{\gamma}^{\,\prime}\right)^2 {\mathcal A} \left(\bar{f}(\bar{\Phi})\right)
		 -6\MP ^2\bar{\gamma}^{\,\prime}\bar{f}^\prime \mathcal{A}^\prime \big] \,.
\end{align}%
Here a prime denotes the differentiation of the corresponding quantity with respect to its argument, for instance 
$\bar{f}^{\,\prime}  \equiv  \mathrm{d}\bar{f}(\bar{\Phi}) / \mathrm{d}\bar{\Phi}$ and $\mathcal{A}^\prime \equiv \mathrm{d} \mathcal{A}(\Phi)/\mathrm{d} \Phi$. 

Picking an explicit form for these four functions designates a theory in a particular conformal frame and sets a specific parametrisation for the scalar field $\Phi$. The expressions for two of the four functions can then be modified to our liking through the above transformations, thereby recasting the original theory in another frame and parametrisation. When $\mathcal{A}$ is identically constant the theory is specified in the Einstein frame, while for a constant $\sigma$ the characterisation is given in the Jordan frame. 
 
As a result of the transformation rules~\eqref{Flanagan.A}-\eqref{Flanagan.B}  the following quantities are invariant under a general composition of conformal rescaling and reparametrisation of the scalar field\footnote{To make the physical implications of these invariants  evident, we  changed the original notation of Ref.~\cite{Jarv:2014hma}, which is recovered through 
$\mathcal{I}_{\mathrm{m}}(\Phi) \equiv\mathcal{I}_{1}(\Phi)$,  
$\mathcal{I}_{\mathcal{V}}(\Phi) \equiv\mathcal{I}_{2}(\Phi)$,
$\sqrt{2} \mathcal{I}_{\mathcal{\phi}}(\Phi) \equiv \mathcal{I}_{3}(\Phi)$ and by suppressing $\MP$ factors since the scalar field in \cite{Jarv:2014hma} is dimensionless. For specific parameterisations it could also be necessary to consider the negative branch of the square root in $\mathcal{I}_{\mathcal{\phi}}(\Phi)$ as explained in~\cite{Jarv:2014hma}.}~\cite{Jarv:2014hma}:
\begin{align}
\label{I.1}
\mathcal{I}_{\mathrm{m}}(\Phi) &\equiv \frac{\mathrm{e}^{2\sigma(\Phi)}}{\mathcal{A}(\Phi)} \,, \\
\label{I.2}
\mathcal{I}_{\mathcal{V}}(\Phi) &\equiv \frac{\mathcal{V}(\Phi)}{\left(\mathcal{A}(\Phi)\right)^2}\,, \\
\label{I.3}
\mathcal{I}_{\mathcal{\phi}}(\Phi) &\equiv \frac{1}{\sqrt{2}} \int \left( \frac{2\mathcal{AB} + 3\left(\mathcal{A}^\prime\right)^2 \MP^{2}}{\mathcal{A}^2} \right)^{\frac{1}{2}}\,\mathrm{d}\Phi    \,.
\end{align}
The quantity $\mathcal{I}_{\mathcal{\phi}}(\Phi)$ provides an invariant description of the scalar degree of freedom and has the corresponding dimension. The integrand in Eq.~\eqref{I.3} can be interpreted as the volume form of the $1$-dimensional space of the scalar field, therefore $\mathcal{I}_{\mathcal{\phi}}(\Phi)$ measures the invariant ``distance'' in such space \cite{Kuusk:2016rso, Kuusk:2015dda}. Constant values of $\mathcal{I}_{\mathcal{\phi}}$ signal that the scalar field is not dynamical, whereas negative values for the expression under the square root in \eqref{I.3} indicate that the theory contains a ghost 
\cite{Jarv:2014hma,Hohmann:2016yfd}.
By inverting the relation \eqref{I.3} and regarding $\mathcal{I}_\phi$ as a new independent degree of freedom in place of $\Phi$, we can write the action \eqref{Flanagan.action} in an invariant fashion \cite{Jarv:2014hma}
\begin{align}
\label{Einstein.frame.action}
S & = \int\mathrm{d}^4x\sqrt{-\hat{g}}\left[ -\frac{\MP ^2}{2}\hat{R}+
\frac{1}{2}\hat{g}^{\mu\nu}\hat{\nabla}_\mu \mathcal{I}_{\mathcal{\phi}}\hat{\nabla}_\nu\mathcal{I}_{\mathcal{\phi}}- \mathcal{I}_{\mathcal{V}}\right]
\nonumber \\
&\qquad
+S_\mathrm{m}\left(\mathcal{I}_{\mathrm{m}} \hat{g}_{\mu\nu},\chi\right) \, ,
\end{align}
where the hatted quantities are functions of the invariant metric $\hat{g}_{\mu\nu}\equiv \mathcal{A}g_{\mu\nu}$. 
The action in Eq.~\eqref{Einstein.frame.action}, which possesses the usual Einstein frame form with respect to the metric $\hat{g}_{\mu\nu}$, clarifies the physical meaning of the remaining invariants \eqref{I.1}-\eqref{I.2}.

$\mathcal{I}_{\mathrm{m}}$ is a dimensionless quantity that characterises the non-minimal coupling in the Jordan frame and, correspondingly, the universal interaction between matter and the scalar field in the Einstein frame. Effectively, $\mathcal{I}_{\mathrm{m}}$ therefore sets the coupling of gravity to the matter fields. For constant $\mathcal{I}_{\mathrm{m}}$ the theory is equivalent to general relativity with a minimally coupled scalar field, otherwise the scalar field participates in mediating the gravitational interaction and is sourced by the trace of the matter energy-momentum tensor \cite{Jarv:2014hma}. The second invariant, $\mathcal{I}_{\mathcal{V}}$, has the dimension of a Lagrangian density and plays the r\^ole of an invariant potential. 

From Eq.~\eqref{Einstein.frame.action} it is also clear that the gravitational aspects of the theory are uniquely specified by only two invariant functions: $\mathcal{I}_{\mathrm{m}}(\mathcal{I}_\phi)$ and $\mathcal{I}_{\mathcal{V}}(\mathcal{I}_\phi)$. The starting action in Eq.~\eqref{Flanagan.action} depends instead on four functions, as it distinguishes between different choices of frame and parametrisation. Consequently, it is not surprising that more than one of these choices could result in the same invariant action \eqref{Einstein.frame.action} once the proposed invariant formalism is applied. This is the case for models that according to Eq.~\eqref{Flanagan.action} differ by the choice of frame and parametrisation, but that are characterised by the same invariants $\mathcal{I}_{\mathrm{m}}(\mathcal{I}_\phi)$ and $\mathcal{I}_{\mathcal{V}}(\mathcal{I}_\phi)$, and therefore for Eq.~\eqref{Einstein.frame.action} by the same invariant action. In this sense, the action \eqref{Einstein.frame.action} provides a solid criterion to establish equivalence classes of gravitational theories and exposes the redundancy implicit in the traditional characterisation provided by Eq.~\eqref{Flanagan.action}. In the following, we will then refer to  theories characterised by the same $\mathcal{I}_{\mathrm{m}}(\mathcal{I}_{\mathcal{\phi}})$ and $\mathcal{I}_{\mathcal{V}}(\mathcal{I}_{\mathcal{\phi}})$ as \emph{equivalent gravitational theories}. 
Clearly, the proposed subdivisions of models in equivalence classes can be further refined in cases where the new scalar degree of freedom possess additional non-gravitational interactions with the matter fields. 

Notice that arbitrarily many different invariants can be defined on the basis of the quantities in Eqs.~\eqref{I.1}-\eqref{I.3}, for instance by forming arbitrary functions of them, $\mathcal{I}_j=f(\mathcal{I}_i)$, by taking a quotient of derivatives, $\mathcal{I}_j \equiv \mathcal{I}'_k/\mathcal{I}'_l \equiv \mathrm{d}\mathcal{I}_k/\mathrm{d}\mathcal{I}_l$, or by integrating, $\mathcal{I}_k=\int \mathcal{I}_j \mathcal{I}'_l \mathrm{d}\Phi$ \cite{Jarv:2014hma}. 
If the physical observables are to be independent of the choice of frame and parametrisation in which a particular theory is specified, we expect that their expression can be given in terms of our invariants. In the next section we  demonstrate the case of inflationary observables.

\section{Inflationary parameters} 
\label{sec:Inflationary parameters}

As Eq.~\eqref{Einstein.frame.action} matches the action in the Einstein frame with respect to the hatted metric, we can easily rephrase the usual expressions for the slow-roll parameters~\cite{Liddle:1994dx,Ade:2015lrj} 
in terms of the invariants in Eqs.~\eqref{I.1}-\eqref{I.3} \cite{Kuusk:2016rso}:
\begin{align}
	\label{epsilon}
	\epsilon &= \frac{\MP ^2}{2}\left(\frac{\mathrm{d} \ln \mathcal{I}_{\mathcal{V}}}{\mathrm{d} \mathcal{I}_{\mathcal{\phi}}}\right)^2\,,\\
	\label{eta}
	\eta &= \frac{\MP ^2}{\mathcal{I}_{\mathcal{V}}}\frac{\mathrm{d}^2 \mathcal{I}_{\mathcal{V}}}{\mathrm{d} \mathcal{I}_{\mathcal{\phi}}^2}\,,\\
	\label{xi}
	\xi^2&=\frac{\MP ^4}{\mathcal{I}_{\mathcal{V}}^2}\frac{\mathrm{d} \mathcal{I}_{\mathcal{V}}}{\mathrm{d} \mathcal{I}_{\mathcal{\phi}}}\frac{\mathrm{d}^3\mathcal{I}_{\mathcal{V}}}{\mathrm{d} \mathcal{I}_{\mathcal{\phi}}^3}\,.
\end{align}
Inflationary observables such as the tensor-to-scalar ratio $r$, the scalar spectral index $n_s$ and the running of the index $\mathrm{d} n_s/(\mathrm{d} \ln k )$ can then be computed in the slow-roll approximation as
\begin{align}
	\label{r}
	r &= 
	8\MP ^2\left(\frac{\mathrm{d} \ln \mathcal{I}_{\mathcal{V}}}{\mathrm{d} \mathcal{I}_{\mathcal{\phi}}}\right)^2 \,,
	\\
	\label{ns}
	n_s &= 
	1 -3\MP ^2 \left(\frac{\mathrm{d} \ln \mathcal{I}_{\mathcal{V}}}{\mathrm{d} \mathcal{I}_{\mathcal{\phi}}}\right)^2+2\frac{\MP ^2}{\mathcal{I}_{\mathcal{V}}}\frac{\mathrm{d}^2 \mathcal{I}_{\mathcal{V}}}{\mathrm{d} \mathcal{I}_{\mathcal{\phi}}^2}  \,,
	\\
	\label{ns.running}
	\frac{\mathrm{d} n_s}{\mathrm{d} \ln k } 
	&= 
	2\MP ^4 \frac{1} 
	{\mathcal{I}_{\mathcal{V}}}\frac{\mathrm{d} \ln \mathcal{I}_{\mathcal{V}}}{\mathrm{d} \mathcal{I}_{\mathcal{\phi}}} \Bigg[4\frac{\mathrm{d} \ln \mathcal{I}_{\mathcal{V}}}{\mathrm{d} \mathcal{I}_{\mathcal{\phi}}}\frac{\mathrm{d}^2 \mathcal{I}_{\mathcal{V}}}{\mathrm{d} \mathcal{I}_{\mathcal{\phi}}^2}
	\nonumber\\ &\quad
	-3 \mathcal{I}_{\mathcal{V}}\left(\frac{\mathrm{d} \ln \mathcal{I}_{\mathcal{V}}}{\mathrm{d} \mathcal{I}_{\mathcal{\phi}}}\right)^3
	-\frac{\mathrm{d}^3\mathcal{I}_{\mathcal{V}}}{\mathrm{d}\, \mathcal{I}_{\mathcal{\phi}}^3}\Bigg] \,,
\end{align}
and the number of $e$-folds of inflation is instead given by
\begin{equation}
	\label{number.of.efolds}
	N(\mathcal{I}_{\mathcal{\phi}}^N) = \frac{1}{\MP ^2}\int\limits_{\mathcal{I}_{\mathcal{\phi}}^\text{end}}^{\mathcal{I}_{\mathcal{\phi}}^N}
	\mathcal{I}_{\mathcal{V}}(\mathcal{I}_{\mathcal{\phi}})\left(\frac{\mathrm{d}\mathcal{I}_{\mathcal{V}}(\mathcal{I}_{\mathcal{\phi}})}{\mathrm{d}\mathcal{I}_{\mathcal{\phi}}}\right)^{-1}\mathrm{d} \mathcal{I}_{\mathcal{\phi}}\,,
\end{equation}
where \(\mathcal{I}_{\phi}^\text{end}\) is the field value at the end of inflation, obtained by solving \(\epsilon(\mathcal{I}_{\phi}^\text{end})=1\).
Finally, we find for the amplitude of the scalar power spectra:
\begin{equation}
	\label{scalar.amplitude}
	A_s = \frac{\mathcal{I}_{\mathcal{V}}}{12\pi^2\,\MP ^6}\left(\frac{\mathrm{d} \ln \mathcal{I}_{\mathcal{V}}}{\mathrm{d} \mathcal{I}_{\mathcal{\phi}}}\right)^{-2} \,.
\end{equation}

The adopted formalism shows that the expressions in Eqs.~\eqref{r}-\eqref{scalar.amplitude} are, as expected, invariant quantities. Furthermore, we find that the analysed observables depend solely on the \emph{invariant potential} $\mathcal{I}_{\mathcal{V}}(\mathcal{I}_{\mathcal{\phi}})$. Therefore, as far as the basic inflationary kinematics is concerned, not only the models which emerge from equivalent gravitational theories yield identical observables, but any class of theories with the same functional form of $\mathcal{I}_{\mathcal{V}}(\mathcal{I}_{\mathcal{\phi}})$ delivers exactly the same phenomenology. 
This insight provides the cornerstone for a classification of phenomenologically equivalent scalar inflation models that we exemplify later in this Letter.

The fact that inflationary observables are independent of $\mathcal{I}_{\mathrm{m}}$ is not surprising: during inflation the dynamics of the scalar field dominates, and the matter part of the action, which involves the invariant non-minimal coupling, is consequently negligible.
$\mathcal{I}_{\mathrm{m}}$ may however play a r\^ole in further distinguishing between inflation models through observables such as the reheating temperature of the Universe, the baryon asymmetry generated, the thermal production of Dark Matter and the non-Gaussianity parameters of inflation~\cite{Martin:2013tda}, which all depend on the couplings of the inflaton to matter. 

\section{Applying the formalism} 
\label{sec:Examples}

The proposed invariant formalism may be applied through the following procedure to any inflationary model that can be cast in the form of the action of Eq.~\eqref{Einstein.frame.action}:
\begin{enumerate}
 \item For a given model of inflation, specified by  an action as in Eq.~\eqref{Flanagan.action}, identify the functions ${\mathcal A}(\Phi)$, ${\mathcal B}(\Phi)$, ${\mathcal V}(\Phi)$ and $\sigma(\Phi)$.
  \item Use Eq.~\eqref{I.3} to compute the invariant $\mathcal{I}_{\mathcal{\phi}}(\Phi)$ and, if possible, invert the relation to obtain $\Phi(\mathcal{I}_{\mathcal{\phi}})$.
  \item Next use $\Phi(\mathcal{I}_{\mathcal{\phi}})$ and Eq.~\eqref{I.2} to calculate the invariant potential $\mathcal{I}_{\mathcal{V}}(\Phi(\mathcal{I}_{\mathcal{\phi}}))=\mathcal{I}_{\mathcal{V}}(\mathcal{I}_{\mathcal{\phi}})$.
 \item With $\mathcal{I}_{\mathcal{V}}(\mathcal{I}_{\mathcal{\phi}})$ and Eq.~\eqref{epsilon} compute $\epsilon=\epsilon(\mathcal{I}_{\mathcal{\phi}})$. Then, provided inflation ends, solve for \(\epsilon(\mathcal{I}_\phi^\text{end})=1\) and integrate Eq.~\eqref{number.of.efolds}
to obtain $N(\mathcal{I}_{\mathcal{\phi}})$. If possible, invert it to obtain $\mathcal{I}_{\mathcal{\phi}}(N)$. 
 \item Once $\mathcal{I}_{\mathcal{\phi}}(N)$ and $\mathcal{I}_{\mathcal{V}}(\mathcal{I}_{\mathcal{\phi}})$ are known, the inflationary observables can be obtained from Eqs.~\eqref{r}-\eqref{scalar.amplitude}.   
\end{enumerate}

We exemplify now the procedure in the case of the Higgs inflation model, specified by~\cite{Bezrukov:2007ep} 
\begin{align}
\label{Higgs.A}
\mathcal{A}(\Phi) &= \frac{ M^{2}+\xi \Phi^{2}}{\MP ^{2}} \,,\\ 
\label{Higgs.B}
\mathcal{B}(\Phi) &= 1 \, ,\\
\label{Higgs.V}
\mathcal{V}(\Phi) &= \frac{\lambda}{4}(\Phi ^2-v^2)^2 \, ,\\
\label{Higgs.alpha}
\mathcal{\sigma}(\Phi) &= 0 \, ,
\end{align}
where $\xi$ is the non-minimal coupling to gravity, $\lambda$ is the Higgs boson self-coupling and $v$ its vacuum expectation value (VEV). We take the latter at its measured value, $v=246$ GeV, and assume that $M\simeq \MP $, $\MP  \ll \xi \Phi$, $\xi \gg 1$. In this regime, $\mathcal{I}_{\mathcal{\phi}}$ Eq.~\eqref{I.3} reduces to
\begin{equation}
	\label{Higgs.I.3}
	\mathcal{I}_{\mathcal{\phi}}(\Phi)=\sqrt{6}\MP \ln\left(\frac{\sqrt{\xi}\Phi}{\MP }\right) \, ,
\end{equation}
where we set $\mathcal{I}_{\mathcal{\phi}}( \MP/\sqrt{\xi} )=0$. Inverting the last expression and using Eq.~\eqref{I.2} results in
\begin{equation}
	\label{Higgs.I.2}
	\mathcal{I}_{\mathcal{V}}(\mathcal{I}_{\mathcal{\phi}})\simeq \frac{\lambda}{4} \frac{\MP ^4}{\xi^2}
	\Biggl[1-\exp \left(-\sqrt{\frac{2}{3}}\frac{\mathcal{I}_{\mathcal{\phi}}}{\MP } \right) \Biggr]^2 \,.
\end{equation}
Next, from Eq.~\eqref{epsilon} and \eqref{eta} we have
\begin{align}
	\label{Higgs.epsilon}
	\epsilon &= \frac{4}{3}\exp\left(-2\sqrt{\frac{2}{3}}\frac{\mathcal{I}_{\mathcal{\phi}}}{\MP }\right)\Biggl[1-\exp\left(-\sqrt{\frac{2}{3}}\frac{\mathcal{I}_{\mathcal{\phi}}}{\MP }\right)\Biggr]^{-2} \,, \\
	\label{Higgs.eta}
	\eta &= \Biggl[2-\exp\left(\sqrt{\frac{2}{3}}\frac{\mathcal{I}_{\mathcal{\phi}}}{\MP }\right)\Biggr]\epsilon \,.
\end{align}
Solving now for $\epsilon(\mathcal{I}_{\mathrm{\phi}}^{\rm end})=1$ yields $\mathcal{I}_{\mathrm{\phi}}^{\rm end}
\simeq 0.94 \MP $. The number of $e$-folds is then given by Eq.~\eqref{number.of.efolds} as
\begin{align}
	\label{Higgs.e.folds}
	N &\simeq \frac{3}{4}\Biggl[\exp\left(\sqrt{\frac{2}{3}}\frac{\mathcal{I}_{\mathcal{\phi}}}{\MP }\right)-\sqrt{\frac{2}{3}}\frac{\mathcal{I}_{\mathcal{\phi}}}{\MP } 
	\Biggr]\, - 1.
\end{align}
By inverting this expression we obtain
 \begin{eqnarray}
\label{Higgs.I.3.in.terms.of.N}
	\mathcal{I}_{\mathcal{\phi}}\simeq \sqrt{\frac{3}{2}}\MP 
	\ln\left[ \frac{4}{3}(N+1) + \ln\left( \frac{4}{3}(N+1) \right) \right] \,,
\end{eqnarray}
and by using Eqs.~\eqref{Higgs.epsilon}, \eqref{Higgs.eta}, \eqref{ns} and Eq.~\eqref{r} we finally have
\begin{align}
	\label{Higgs.n.s}
	n_{s}&\simeq  1-\frac{2}{N+1}-\frac{9-3\ln\Bigl[\frac{4}{3}(N+1)\Bigr]}{2(N+1)^2} \,,\\
	\label{Higgs.r}
	r&\simeq \frac{12}{(N+1)^2}\Biggl\{ 1 + \frac{3-3\ln\Bigl[\frac{4}{3}(N+1)\Bigr]}{2(N+1)}\Biggr\} \,.
\end{align}

These formulae correctly reproduce the results of the original calculations in Einstein~\cite{Bezrukov:2007ep} and Jordan frame~\cite{Tsujikawa:2004my,vandeBruck:2015gjd} in the leading order.

\begin{table*}[t]
\caption{The first class of inflationary models we consider encompasses the models of quadratic and Coleman-Weinberg inflation in induced gravity.}
\begin{center}
\begin{ruledtabular}
\begin{tabular*}{\textwidth}{cccccccc}
  & $\mathcal{A}$ & $\mathcal{B}$ & $\mathcal{V}$ & $\mathcal{\sigma}$ & $\mathcal{I}_{\mathrm{m}}$ & $\mathcal{I}_{\mathcal{V}}$
  & $\mathcal{I}_{\mathrm{\phi}}$ 
  \\
\hline \\[-0.5em]
 Quadratic & $1$ & $1$ & $\frac{1}{2} M^{2} \Phi^{2}$ & $0$ & $1$ & $\frac{1}{2} M^{2} \Phi^{2}$
  & $\Phi$
  \\
\parbox[t]{5cm}{\centering Coleman-Weinberg inflation\\ in induced gravity} & $\xi \frac{\Phi^{2}}{\MP^{2}}$ & $1$ & $\frac{1}{4} K (\ln^{2} \frac{\Phi}{v_{\Phi}}) \Phi^{4}$ & $0$ & $\frac{1}{\xi} \frac{\MP^{2}}{\Phi^{2}}$ & $ \frac{1}{4} K (\ln^{2} \frac{\Phi}{v_{\Phi}}) \frac{\MP^{4}}{\xi^{2}} $
  & $\MP \sqrt{\frac{1 + 6 \xi}{\xi}} \ln \frac{\Phi}{v_{\Phi}}$\\
\end{tabular*} 
\end{ruledtabular}
\end{center}
\label{tab:first:class}
\end{table*}%

\setlength{\tabcolsep}{5pt}
\begin{table*}[t]
\caption{The second class of inflationary models we identify encompasses $\alpha-\beta$ model (M1), the E-type $\alpha$-attractors (M2) and special $\xi$-attractors (M3).}
\begin{ruledtabular}
\begin{tabular*}{\textwidth}{cccccccc}
  & $\mathcal{A}$ & $\mathcal{B}$ & $\mathcal{V}$ & $\mathcal{\sigma}$ 
  & $\mathcal{I}_{\mathrm{m}}$ & $\mathcal{I}_{\mathcal{V}}$ & $\mathcal{I}_{\mathrm{\phi}}$ 
  \\ 
  \hline \\[-0.9em]
  M1 & $1$ & $1$ & $M^{4} \left( 1 - e^{-\sqrt{\frac{2}{3\alpha}} \frac{\Phi}{\MP}} \right)^{2}$ & $0$ 
  & $1$ & $M^{4} \left( 1 - e^{-\sqrt{\frac{2}{3 \alpha}} \frac{\Phi}{\MP}} \right)^{2}$
  & $\Phi$
  \\
  M2 & $1$ & $\frac{3 \alpha}{2} \frac{\MP^{2}}{\Phi^{2}}$ & $M^{4} \left(1 - \frac{\Phi}{\MP} \right)^{2}$ & $0$ 
  & $1$ & $M^{4} \left(1 - \frac{\Phi}{\MP} \right)^{2}$ & $\
 -\sqrt{\frac{ 3\alpha}{2}} \MP \ln \frac{\Phi}{\MP}$
  \\
 M3 & $\frac{\MP^{2} + \xi \Phi^{2}}{\MP^{2}}$ & $\frac{\xi \Phi^{2}}{\MP^{2} + \xi \Phi^{2}}$ & $\lambda \xi^{2} \Phi^{4}$
  & $0$
  & $\frac{\MP^{2}}{\MP^{2} + \xi \Phi^{2}}$ & $\frac{\lambda \xi^{2} \MP^{4} \Phi^{4}}{(\MP^{2} + \xi \Phi^{2})^{2}}$ & $\frac{\sqrt{\xi (1 + 6 \xi)}}{2 \xi} \MP\ln \left( 1 + \frac{\xi \Phi^{2}}{\MP^{2}} \right)$
  \\
\end{tabular*} 
\end{ruledtabular}
\label{tab:second:class}
\end{table*}%

\section{Identifying equivalent models}

The invariant formalism proposed in this Letter allows to identify classes of models which yield identical inflationary phenomenologies in a straightforward way.
In spite of the different Lagrangians, models that yield the same invariant potential $\mathcal{I}_{\mathcal{V}}(\mathcal{I}_{\mathcal{\phi}})$ necessarily result in the same ranges of the relevant inflationary observables.  
By using the procedure delineated above, it is therefore possible to categorise the known models of inflation in equivalence classes which, unequivocally, correspond to phenomenologically different frameworks.\footnote{Within each class, we can construct arbitrarily many equivalent models by simply taking a different form for $\Phi(\mathcal{I}_{\mathcal{\phi}})$.}
We give an example of the power of this formalism by identifying two classes of phenomenologically equivalent models.
 
\subsection{Quadratic inflation and Coleman-Weinberg inflation in induced gravity} 
\label{sub:A first class}

As a first example we consider here the cases of quadratic and Coleman-Weinberg inflation in induced gravity, presented in detail in \cite{Linde:1983gd} and \cite{Kannike:2015apa}, respectively. 

According to our procedure, the models are specified in the first four columns of table~\ref{tab:first:class}, whereas the last three columns present the expressions for the invariants $\mathcal{I}_{\mathrm{m}}$, $\mathcal{I}_{\mathcal{V}}$ and $\mathcal{I}_{\mathcal{\phi}}$ that we obtained in these cases. The constant $K$ and the logarithm squared in the induced gravity model arise from the first non-zero term of the Taylor expansion of the running scalar coupling $\lambda_{\Phi}(\Phi)$ \cite{Kannike:2015apa}, given by
\begin{equation}
  \lambda_{\Phi}(\Phi) \approx \frac{1}{2!} \beta'_{\lambda_{\Phi}} \ln^{2} \frac{\Phi}{v_{\Phi}} 
  \equiv  K \ln^{2} \frac{\Phi}{v_{\Phi}}\,,
\end{equation}
where $ v_{\Phi} = \MP/\sqrt\xi$ and $\beta_{\lambda_{\Phi}}$ is the $\beta$-function for the scalar self-coupling $\lambda_{\Phi}$.

It is straightforward to show through a direct calculation that both these models share the same invariant potential
\begin{equation}
  \mathcal{I}_{\mathcal{V}}(\mathcal{I}_{\mathrm{\phi}}) = \frac{1}{2} M^{2} \mathcal{I}_{\mathrm{\phi}}^{2}\,,
\end{equation}
identifying
\begin{equation}
  M^2 = \frac{K}{2\xi (1+6\xi)} \MP^2.
\end{equation}

Therefore, for Eqs.~\eqref{r}-\eqref{scalar.amplitude}, the two models result undoubtably in the same phenomenology despite being distinct gravitational theories characterised by different $\mathcal{I}_{\mathrm m}$.

\subsection{A second example: $E$-type models} 
\label{sub:Another example}

We consider now the following models of inflation: the generalisation of the Starobinsky potential ($\alpha-\beta$ model) \cite{Ferrara:2013rsa} (M1), the E-type $\alpha$-attractor \cite{Kallosh:2013yoa} (M2) and the special $\xi$-attractor \cite{Galante:2014ifa} (M3). As before, table~\ref{tab:second:class} presents
the specifications of these models in terms of the functions ${\mathcal A}(\Phi)$, ${\mathcal B}(\Phi)$, ${\mathcal V}(\Phi)$ and $\sigma(\Phi)$, showing as well the corresponding expressions for the invariants. 

A straightforward calculation shows that the models yield the same invariant potential
\begin{equation}
  \mathcal{I}_{\mathcal{V}}(\mathcal{I}_{\mathrm{\phi}}) =M^4 \left(1 - e^{-\sqrt{\frac{2}{3\alpha}} \frac{\mathcal{I}_{\mathrm{\phi}}}{\MP} } \right)^2\,,
\end{equation}
identifying
\begin{equation}
  \alpha = 1 + \frac{1}{6 \xi}\,,\qquad M^4=\lambda \MP^4\,,
\end{equation}
and therefore to the same phenomenology. The fact that these models result in same inflation features has been previously noticed in literature \cite{Galante:2014ifa} and is made manifest with our formalism.

\section{New directions in model building}
\label{sec:More models}

Another perk of the proposed invariant formulation is that it allows the study of models detailed in invariant potentials which are not elementary functions of $\mathcal{I}_\phi$. Consider for instance a scenario specified by $\mathcal{A}=1$, $\mathcal{B}=e^{-\frac{\Phi^2}{\MP^2}}$, $\mathcal{V}=M^4 e^{-\frac{b\Phi}{\MP}}$, resulting in
\begin{equation}
\mathcal{I}_\phi = \sqrt{\frac{\pi}{2}} \MP \, \mathrm{Erf}\left(   \frac{1}{\sqrt{2}}\frac{\Phi}{\MP} \right) \,,
\end{equation}
where $\mathrm{Erf}$ is the ``error function'' usually appearing in statistics. As its inverse function, $\mathrm{InvErf}$, is also known, we obtain 
\begin{equation}
\mathcal{I}_\mathcal{V} (\mathcal{I}_\phi) = M^4 \exp\left(  -\sqrt{2}b \, \mathrm{InvErf} \sqrt{\frac{2}{\pi}} \frac{\mathcal{I}_\phi}{\MP} \right) \,.
\end{equation}
By computing the slow-roll parameters via Eq.~\eqref{epsilon} and \eqref{eta} 
\begin{eqnarray}
\epsilon &=& \frac{b^2}{2} \, e^{2 \left(\mathrm{InvErf}\sqrt{\frac{2}{\pi}} \frac{\mathcal{I}_\phi}{\MP} \right)^2 } \,, \\
\eta &=& b \left( b - \sqrt{2} \, \mathrm{InvErf}\sqrt{\frac{2}{\pi}} \frac{\mathcal{I}_\phi}{\MP} \right) e^{2 \left( \mathrm{InvErf}\sqrt{\frac{2}{\pi}}\frac{\mathcal{I}_\phi}{\MP} \right)^2} \,.
\end{eqnarray}
we find that for $0<b\ll1$ this atypical model yields a sufficiently long inflationary era with properties allowed by the latest measurements of $r$ and $n_s$ at a confidence level of about $95\%$.

This basic example proves that inflationary models specified by elementary functions that supposedly arise from fundamental physics can lead to invariant potentials given in terms of special functions. The proposed formalism is suited to study such atypical scenarios, the phenomenology of which lies outside the boundaries of the current compendia like Ref.~\cite{Martin:2013tda}.

\section{Conclusions} 
\label{sec:Conclusions}

The main objective of the present Letter was to identify the origin of the redundancy in the current description of inflation and propose an alternative and clearer categorisation of the viable inflationary scenarios. 

To this purpose, by adopting a formalism in which slow-roll parameters and inflationary observables can be expressed in a frame and parametrisation invariant fashion, we demonstrated that the phenomenology of every inflation model is solely regulated by the so-called invariant potential. As a result, it is obvious that models characterised by identical invariant potentials lead to the same physical consequences, in spite of the different starting Lagrangians.

After detailing how to recast a general model of inflation in the proposed  formalism, we exemplified the procedure in the case of the Higgs inflation. With the invariant formalism at hand, we then demonstrated the physical equivalence of different inflationary scenarios proposed in literature, posing the basis for the sought categorisation of viable inflation models and for a better understanding of the connected dynamics.

In this regard, we proved that the standard quadratic inflation model and the more recent induced Coleman-Weinberg scenario give rise to twin phenomenologies delineating a first class of equivalent theories. Likewise, we showed that $\alpha-\beta$ models, E-type $\alpha$-attractors and special $\xi$-attractors fall into a second equivalence class.  

On top of that, we showed how the proposed formalism can be employed to study the phenomenology of viable inflationary models encoded in invariant potential specified by special functions. These scenarios lie outside of the boundaries of the current categorisation of inflationary frameworks and, therefore, represent a new possible direction of model building.  

It is our hope that the methodology proposed in this Letter will become the new language for the characterisation of viable inflationary scenarios, paving the way toward a deeper understanding of the dynamics of inflation itself. 


\begin{acknowledgments}
\section*{Acknowledgments}
The authors thank Ott Vilson for useful discussions.
This work was supported by the Estonian Research Council grants IUT23-6, IUT02-27, PUTJD110, PUT790, PUT799, PUT1026, and by EU through the ERDF CoE program with the grant TK133.
\end{acknowledgments}


%

\end{document}